\documentclass[10pt,journal,onecolumn]{IEEEtran}

\usepackage[utf8]{inputenc} 
\usepackage[T1]{fontenc}
\usepackage{url}
\usepackage{ifthen}
\usepackage[cmex10]{amsmath}

\usepackage{multicol}

\usepackage[margin=1in, top=1.2in]{geometry}

\usepackage{hyphenat}
\usepackage{fancyhdr}
\usepackage{graphicx}
\usepackage{color}
\usepackage{float}

\usepackage[allcolors=red, colorlinks=true, urlcolor=blue]{hyperref}

\usepackage{enumerate}

\usepackage{dsfont}
\usepackage{bbm}

\usepackage{amsmath}   
\usepackage{mathtools}
\usepackage{amssymb}
\usepackage{amsbsy}
\usepackage{amstext}
\usepackage{amscd}
\usepackage{amsopn}
\usepackage{amsthm}
\usepackage{amsfonts}
\usepackage{xfrac}
\usepackage{mathabx}
\usepackage{framed}
\usepackage{adjustbox}

\usepackage{setspace}

\usepackage[numbers,sort&compress]{natbib}
\usepackage{bibentry}

\newcommand{\RR}{\mathds{R}}

\newcommand{\pr}[1]{\mathrm{Pr}\left\{#1\right\}} 
\newcommand{\evv}[2]{\mathds{E}_{#1}\left\{#2\right\}}  
\newcommand{\myvarr}[2]{\mathds{V}\!\mathrm{ar}_{#1}\left\{#2\right\}}  

\newcommand{\sd}{\mathbf{SD}}
\newcommand{\myop}[1]{\mathsf{#1}}

\newcommand{\eps}{\epsilon}
\renewcommand{\varepsilon}{\epsilon}

\newcommand{\mc}[1]{\mathcal{#1}}

\providecommand{\eqref}[1]{(\ref{#1})}
\newcommand{\vect}[1]{\boldsymbol{\mathrm{#1}}}
\newcommand{\mymtrx}[1]{\boldsymbol{\mathsf{#1}}}

\newcommand{\remove}[1]{}

\makeatletter
\def\th@plain{%
	\thm@notefont{}
	\itshape 
}
\def\th@definition{%
	\thm@notefont{}
	\normalfont 
}
\makeatother

\newtheorem{theorem}{Theorem}

\newtheorem{lemma}[theorem]{Lemma}

\newtheorem{remark}{Remark}
\newtheorem{corollary}{Corollary}[theorem]

\theoremstyle{definition}
\newtheorem{definition}{Definition}
\theoremstyle{definition}

\newlength{\eqboxstorage}

\begin{document}
	\title{A Channel Model of Transceivers for Multiterminal Secret Key Agreement}
	
	\author{
		\IEEEauthorblockN{Alireza Poostindouz, and Reihaneh Safavi-Naini \\}
		\IEEEauthorblockA{University of Calgary\\
			Calgary, AB, Canada}
	}

	\maketitle

	\begin{abstract}
		
		Information theoretic secret key agreement is impossible without making initial assumptions. 
		One type of initial assumption is correlated random variables that are generated by using a noisy channel that connects the terminals.  
		Terminals  use  the correlated random variables and communication over a reliable public channel to arrive at a   
		shared secret key.  
		Previous channel models assume that each terminal either controls  one input to the channel, or receives
		one 
		output variable of the   
		channel. 
		In this paper, we propose a new channel model of transceivers where each terminal  simultaneously controls an input variable and observes an output variable of the (noisy) channel. 
		We give upper and lower bounds for the secret key capacity (i.e., highest achievable key rate) of this transceiver model, and prove the  secret key capacity under the conditions that the public communication is noninteractive and input variables of the noisy channel are independent. 
	\end{abstract}
	
	\section{Introduction}\label{sec:Intro}

	Secret key agreement (SKA) protocols are  an essential component of cryptosystems.  
	The information theoretic treatment of the problem of two-party SKA  was first considered 
	in  \cite{Maurer1993},  and  independently  in~\cite{Ahlswede1993}.
	These results were later extended to multiterminal scenarios \cite{Csiszar2004a,Csiszar2008,Csiszar2013}. 
	In this paper we consider the multiterminal SKA problem.

	Information theoretic secure 
	SKA is impossible without initial assumptions. 
	A commonly used assumption is the existence of correlated random variables at the terminals. 
	In the {\em channel model} of 
	SKA, a noisy channel is used to generate the correlated variables. 
	In the {\em  multiterminal channel model } for  SKA, there is a set of $m$ terminals, 
	denoted by $\mc M=\{1,\ldots,m\}$, and 
	the goal of the SKA protocol is to establish a shared secret key among a designated subset $\mc A\subseteq \mc M$ of terminals. 
	There exists 
	a noisy DMC (discrete memoryless channel) which will be used for generating correlation among 
	terminals. 
	Terminals also can send public messages over a public, noiseless, and authenticated channel. Each public message sent by a terminal is publicly  accessible (to all terminals and the adversary).

	The adversary, Eve, has access to all the public messages, and additionally  it may have access 
	to some of the input or output symbols of the channel by compromising a  
	subset  $\mc D\subseteq \mc M\setminus\mc A$ of terminals.  
	At the end of the SKA protocol, terminals in 
	$\mc A$  will obtain the same secret key, such that Eve has no information about it.  
	
	The key capacity of a model is defined with respect to the adversary model, and is the highest achievable secret key rate of SKA under
	the specific  
	adversary model 
	\cite{Csiszar2004a,Csiszar2013}. 
	In this paper, we focus on two notions of capacity. 
	The \textbf{secret key (SK) capacity}, is the key capacity when the adversary has only access to 
	all public messages, and the \textbf{private key (PK) capacity} is the key capacity when the adversary also knows 
	the random variables of a subset $\mc D\subseteq \mc M\setminus \mc A$ of  terminals.

	All existing channel models assume that a terminal either controls an input,  or have access to an output symbol, of the underlying DMC. 
	In this paper, we introduce a new   
	channel model   
	that  we call the \emph{``channel model of transceivers''}, in which
	each terminal   
	provides input to, and receives output from, the channel. 
	The variable associated with a  
	terminal $j\in\mc M$, is of the form $X_j=(T_j,Y_j)$, where $T_j$'s are input variables and $Y_j$'s are output variables of the DMC. 
	This model has  
	the channel models of \cite{Csiszar2008} and \cite{Csiszar2013} as special cases. 
	
	The results of \cite{Csiszar2008,Csiszar2013} are not directly applicable to the  
	channel model of transceivers  because of the dual role of each terminal; %
	however,   
	we prove general lower and upper bounds for the SK and PK capacities of our proposed model using proof ideas from those works. These bounds are not tight in general, and we leave the problem of finding tighter general bounds for future works. 
	We also consider the transceiver model under the constraints that 
	input variables of the transceiver DMC are generated independently, and
	the public communication is noninteractive. 
	For this case, we prove the tightness of the above mentioned bounds and give the noninteractive SK capacity. 
	Finally we give simpler expressions for  the SK capacity of a
	the special channel model which we call the Polytree-PIN.

	
	Our work raises many interesting questions for future work, including finding tighter bounds for the SK and PK capacities, and investigation of interactive protocols for achieving the key capacity of wiretapped Polytree-PINs.

	
		\subsection{Related Works}
		Different channel models differ in their specifications on how terminals control input, and access the output of the noisy DMC, or how terminals are allowed to use the public channel. 
		Single-input multiterminal DMC's where considered  in the channel model of \cite{Csiszar2008} 
		where   $\mc A\subseteq \mc M$, and   
		all terminals are allowed to send public messages.  
		In the single-input multiterminal channel model of \cite{Gohari2010a} however, where $\mc A=\mc M$,  only a subset of terminals $\mc U\subseteq \mc M$ participate in the public communication while the rest of the terminals are silent  (i.e., not sending public messages). 
		An important generalization of the model in \cite{Csiszar2008} is the \emph{multiaccess} channel model of \cite{Csiszar2013}, in which    
		all terminals are allowed to send public messages while  a subset of terminals are supplying input to the DMC and the remaining terminals (which is  a disjoint subset from the first subset) are receiving channel outputs.

		The known results on multiterminal SK and PK capacity are, SK and PK capacities in \cite{Csiszar2008}, and upper and lower bounds of \cite{Gohari2010a} for the single-input models. 
		For the {multiaccess} channel model, the SK and PK capacities are not known in general. 
		Upper bounds and lower bounds on the SK and PK capacities of the {multiaccess} model were given in \cite{Csiszar2013}, 
		where the lower bounds are based on source emulation approach. 
		We use this approach to derive a lower bound for our proposed model (see Section~\ref{sec:lower}).  
		The SK capacity is proved for the \emph{symmetric multiaccess channel}\footnote{A two input single out put channel is called symmetric if the conditional distribution of the channel satisfies $P_{X_3|X_1 X_2} = P_{X_3|X_2 X_1}$. } with single output under the constraint that input terminals are silent \cite{Tyagi2013d}. 
		It was showed that this SK capacity is achievable by an interactive SKA protocol.

	\section{Background}\label{sec:background}

	\subsection{Notation}\label{sec:notation}

	We restrict ourselves to  probability distributions over finite alphabets.
	We reserve upper-case letters  
	for random variables (RVs) and lower-case letters  
	to denote their realizations. Upper-case calligraphic letters (e.g., $\mc{M}$, $\mc A$, etc.\!)  
	denote sets,  
	and for any natural number $m$ we define $[m]:=\{1,\ldots,m\}$. Let ${\mc{M}=[m]}$, then ${X_{\mc{M}}:=(X_1,\ldots,X_m)}$ and ${X_{\mc{A}} = (X_j|~\forall j\in \mc{A})}$ for any $\mc{A}\subseteq \mc{M}$. 
	For an arbitrary real vector $R_{\mc{M}}=(R_1,\ldots,R_m)\in\RR^m$ and for any $\mc{A}\subseteq [m]$ we define $R_{\mc{A}}=(R_j|~\forall j\in \mc{A})$, and $\myop{sum}(R_{\mc{A}}):=\sum_{j\in \mc{A}} R_j$.

	For a given alphabet $\mc W$, let $W$ be a random variable over $\mc W$ and let $P_W$ be the probability distribution of $W$. 
	We then define expectation and variance of any function $f(W)$ as $\evv{P}{f(W)} = \sum_{w\in\mc W} P(w)f(w),$ and $
	\myvarr{P}{f(W)} = \sum_{w\in\mc W} P(w)(f(w) - \evv{P}{f(w)})^2$. 
	The statistical distance between two distributions $P_X$ and $Q_X$ defined over the same alphabet $\mc X$ is also given by
	\begin{equation*}
	\sd(P,Q) = \frac{1}{2} \sum_{x\in\mc X} |P(x) - Q(x)| = \frac{1}{2} \evv{P}{\left\vert 1-\frac{Q}{P} \right\vert}.
	\end{equation*}

	\subsection{The Multiterminal Source Model}
	\label{sec:source_model}

	The general multiterminal source model,  was introduced in \cite{Csiszar2004a}. 
	In this model, there is a set of $m$ terminals 
	denoted by $\mc{M} =[m]=\{1,\ldots, m \}$. 
	Each terminal $j\in[m]$ has access to a random variable $X_j$. 
	Let $X_{\mc{M}}=(X_1, \ldots, X_m)$ denote the set of all variables accessible to all terminals.  
	After  $n$ IID sampling from $X_{\mc{M}}$, terminals use  
	a public channel, that is reliable and 
	authenticated, for a finite number of rounds. 
	A  message that is sent by  terminal $j$ is  
	a function of  the terminal's IID samples (observations)  $X_j^n$, local randomness, and previous public messages.
	We denote by $\vect{F}$ the set of all messages sent over the public channel.

	Let $\mc A\subseteq \mc M$ be the set of terminals  that want to establish 
	a shared secret key, which need not to be  
	fully concealed from the helper terminals in $\mc{A}^c =\mc M\setminus\mc A$.  
	Eve samples from the side information variable $Z$,  which is correlated with 
	$X_\mc{M}$,  and has full read  access  to  
	public messages $\vect{F}$. 
	Eve is a passive adversary, that listens to the public communication only. 
	We denote a source model by its joint distribution $P_{ZX_\mc{M}}$, which is publicly known.   
	A secret key $K$ for terminals in   $\mc{A}$  is considered to be secure against Eve, 
	if it satisfies the reliability and secrecy conditions as defined below.

	\begin{definition}\label{def:SK-Key}
		Consider a set of $m$ terminals $\mc M$, where 
		$\mc A\subseteq \mc M$ denotes the  set of terminals that will share a 
		key  $K$ with alphabet $\mc{K}$. Let $Z^n$ denote  
		Eve's  side information   
		about $X_{\mc M}^n$. 
		The key $K$ is an $(\epsilon,\sigma)$-Secret Key (in short $(\epsilon,\sigma)$-SK) for $\mc A$,  
		if there exists an SKA protocol with public communication $\vect F$, and output RVs $\{K_j\}_{j\in\mc{A}}$ for each terminal, such that 
		\begin{align}\label{eq:rel}
		\text{(reliability)}\quad& \pr{K_j=K}\geq 1-\epsilon, \quad \forall j\in\mc A,\\ \label{eq:secrecy}
		\text{(secrecy)}\quad& \sd\left((K,{\vect{F}} ,Z^n);(U,{\vect{F}},Z^n)\right) \leq \sigma,
		\end{align}
		where $\sd$ denotes the statistical distance and $U$ is the uniform probability distribution over alphabet $\mc{K}$.
	\end{definition}

	For a given source model $P_{ZX_\mc{M}}$ the maximum secret key rate ($(1/n)\log{|\mc K|}$) is called the source model \emph{wiretap secret key (WSK)} capacity. 
	If we restrict Eve's side information to be $Z=\text{constant}$, then the maximum secret key rate is called the source model \emph{secret key (SK)} capacity. 
	If Eve's side information is of the form $Z=X_{\mc D}$, 
	where $\mc D\subseteq \mc A^c$ is the set of compromised terminals, 
	the  maximum secret key rate is called the source model \emph{private key (PK)} capacity. 
	Single-letter characterizations of SK and PK capacities for the general multiterminal source model are given in \cite[Theorems 1 and 2]{Csiszar2004a}. We review these results later in Section \ref{sec:lower}.

	\subsection{The Multiaccess Channel Model}\label{sec:multiaccess}
	
	The multiaccess source model,  was introduced in \cite{Csiszar2013}. 
	Our work generalizes this model, that we will review below. 
	In the multiaccess model, there is a set of $m$ terminals 
	denoted by $\mc{M} =[m]=\{1,\ldots, m \}$. 
	A subset of terminals $\mc I = [k] = \{1,\ldots, k\}$ are called \emph{input terminals}, 
	the rest of terminals in $\mc M\setminus \mc I$ are called \emph{output terminals}. 
	There exists a secure noisy DMC between input terminals and output terminals. 
	Input terminals supply input symbols $X_j~j\in\mc I$ to the DMC, 
	and output terminals observe respective output symbols of the DMC. 
	The underlying noisy DMC is called a multiaccess channel and is denoted by $W=(\mc X_{{\mc{I}}}, P_{X_{{\mc{M} \setminus \mc I}}| X_{{\mc{I}}}}, \mc X_{{\mc{M} \setminus \mc I}})$, where
	\begin{equation*}  
	P_{X_{{\mc{M} \setminus \mc I}}| X_{{\mc{I}}}} : \mc X_1 \times \cdots \times \mc X_k \mapsto \mc X_{k+1} \times \cdots \times \mc X_m. 
	\end{equation*}

	In the basic multiaccess model, Eve does not have any information about transmission over the DMC. 
	Terminals are allowed to use the underlying multiaccess DMC in $n$ rounds, where each round of symbol transmission over the DMC  is followed  by rounds of public channel discussion, and all public messages are accessible to all terminals and Eve.
	The SK and PK capacities for multiaccess channel model are defined similar to the source model.

	General upper bounds and lower bounds were proved in \cite{Csiszar2013} for the SK and PK capacities of the multiaccess channel model.  
	The lower bounds are based on SKA protocols  in \cite{Csiszar2004a} and use a technique called \emph{source emulation} that we will also use in our proofs (see Section \ref{sec:lower}).

	In the next section, we introduce the channel model of transceivers.

	\section{A General Channel Model of Transceivers}\label{sec:model}

	\subsection{The Model}
	\label{sec:defs_model}

	Consider a set of $m$ terminals denoted by $\mc{M} =[m]:=\{1,\ldots, m \}$. 
	The goal of an SKA protocol is for terminals in $\mc M$ to cooperate (using the public communication) so that terminals in a subset $\mc A\subseteq \mc M$ can establish a shared secret key $K$. 
	Terminals in $\mc A^c = \mc M\setminus\mc A$ are called \emph{helper terminals}. 
	The key $K$ is not required to be concealed from the helper terminals. 
	A terminal $j\in{\mc{M}}$ has access to samples of a random variable (RV) denoted by $X_j$. 
	The variables of terminals are correlated.
	Let $X_{\mc{M}}=(X_1, \ldots, X_m)$  
	denote the set of  
	these variables.  
	All terminals have access to a public, reliable, and authenticated channel. A public message sent by a terminal $j$ will be received by all terminals and everyone else, including the passive adversary Eve, who will not interfere with the public communication. 
	Eve may also have access to side information $Z$ which is correlated with $X_{\mc M}$.

	There exists an underlying DMC (discrete memoryless channel) which will be used for generating the correlation in $X_{\mc M}$. 
	For each transmission over the channel, all terminals provide input to the noisy channel \emph{and} receive output from it; i.e., we assume   
	a set of ``\emph{transceivers}''. 
	Each terminal $j$ has two RVs,  $T_j$ which is an input variable to the  
	DMC, and $Y_j$ which is an output variable of the DMC, and so  
	the RV associated with each terminal $j$ is  given by $X_j=(T_j,Y_j)$, where  $\mc X_j=\mc T_j\times \mc Y_j$. The underlying multi-input multi-output DMC is denoted by $W=(\mc T_{{\mc{M}}}, P_{ZY_{{\mc{M}}}| T_{{\mc{M}}}}, \mc Y_{{\mc{M}}}\times \mc Z)$, where
	\begin{equation}\label{eq:channelmodel_TxRx}
	P_{ZY_{{\mc{M}}}| T_{{\mc{M}}}} : \mc T_1 \times \cdots \times \mc T_m \mapsto \mc Y_1 \times \cdots \times \mc Y_m \times \mc Z
	\end{equation} is the transition matrix defined over the finite input alphabet $\mc T_1 \times \cdots \times \mc T_m$ and finite output alphabet $ \mc Y_1 \times \cdots \times \mc Y_m \times \mc Z$. 
	
	An SKA protocol consists of $n$ rounds, 
	where each round consists of one invocation of the noisy channel, 
	followed by public communication by terminals in $\mc M$ over the public  channel. 
	Let $\vect F^{t}$ denote the random variable representing all public messages of the $m$ terminals in round $1\leq t\leq n$, 
	and let $\vect F = (\vect F^1,\ldots,\vect F^n)$ denote the entire public communication during the SKA protocol. 
	Each public message of terminal $j$ in round $1\leq t\leq n$
	is 
	a function of  all  
	previous samples  $X_{j1},X_{j2},\ldots,X_{jt}$, its local randomness,   
	public  messages  
	of the previous rounds ($\vect{F}^{1},\vect{F}^{2},\ldots,\vect{F}^{t-1}$),  and previous public message sent in round $t$.    
	Importantly, terminals have control over  $T_j$'s (input symbols) 
	and can choose them depending on previous input symbols 
	and the previous  public messages that are transferred in each previous round of the SKA  protocol. 
	More specifically,  each input symbol $T_{jt}$ of round $t\geq 2$ may 
	depend on  previous public discussions $\vect{F}^{1},\vect{F}^{2},\ldots,\vect{F}^{t-1}$, and previous samples  $X_{j1},X_{j2},\ldots,X_{j(t-1)}$. Eve has access to  all public messages, $\vect F$, and side information $Z^n$.

	\begin{figure}[t]
		\centering
		\includegraphics[width=0.65\linewidth]{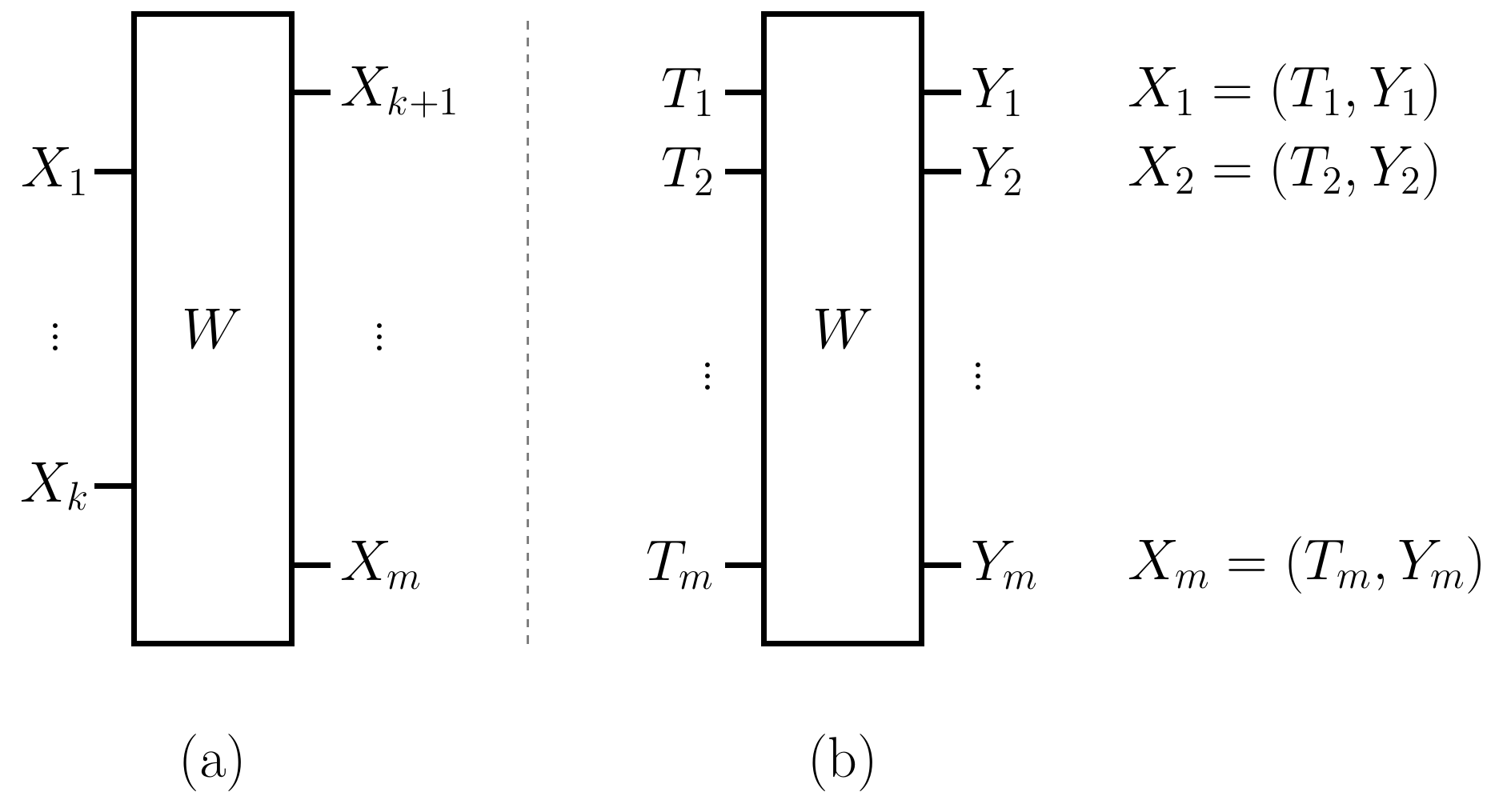}
		\caption{(a) The multiaccess channel model of Ref.~\cite{Csiszar2013} where $P_{ X_{[m]\setminus [k]} | X_{[k]} }$ is the probability transition matrix of DMC $W$. (b) Our proposed general channel model, where $P_{Y_{[m]}| T_{[m]}}$ denotes the transition matrix and for each transceiver terminal $j$; we have $X_j=(T_j,Y_j)$. Eve's side information is assumed to be $Z=\text{constant}$ for both examples here.}
		\label{fig:2channelmodels}
	\end{figure}

	\begin{remark}[Relation with multiaccess channel model]\label{rem:Tx_gen_multiaccess}
	We note that the multiaccess channel model of \cite{Csiszar2013} is a special case of  the 
	channel model of transceivers
	by taking $Z=\text{constant}$, $X_j=T_j~\forall j\in[k]=\{1,\ldots,k\}$, and $X_j=Y_j~\forall j\in[m]\setminus[k]$. 
	See Figure~\ref{fig:2channelmodels} for a pictorial comparison between the 
	channel model  of transceivers,  and the multiaccess  channel model of \cite{Csiszar2013}. The channel model of \cite{Csiszar2008} is a special case of the multiaccess model of \cite{Csiszar2013} for $k=1$, and so a special case of our proposed model.
	\end{remark}

	\subsection{A Unified Definition of Key Capacity}\label{sec:other_defs}

		\vspace{1mm}
		\noindent{\textbf{Noninteractive public communication.}} The use of public communication in   
		channel model is called \emph{``noninteractive''} \cite{Csiszar2013} if during the SKA protocol, terminals each send a single message over the public  channel, and this public communication is after all symbol transmissions over the DMC. In this case, $\vect F=\vect F^n = (F_1,\ldots , F_m)$, where $F_j$ denotes the public message of terminal $j$ which is only a function of $X_j^n$.

		\begin{definition}[Key Capacity]\label{def:SK-Capacity}
			Consider multiterminal SKA for a subset  $\mc A\subseteq \mc M$. Let $Z^n$ denote 
			Eve's  side information  
			about $X_{\mc M}^n$.
			For a given (source or channel) model $Q$, where $Q$ is either the IID source distribution $P_{ZX_{\mc M}}$ or the conditional distribution of the underlying DMC, a real number $R\geq 0$ is an achievable SK rate if there exists an SKA protocol that for every $n$ establishes an $(\eps_n,\sigma_n)-$SK $K\in\mc K$ where $\lim_{n\to\infty} \eps_n = \lim_{n\to\infty} \sigma_n = 0$, and $\lim_{n\to\infty} \frac{1}{n}\log|\mc K| = R$. The supremum of all achievable SK rates
			is called the key capacity of given model $Q$. 
		\end{definition}

	\vspace{1mm}
	\noindent{\textbf{SK, PK, and WSK Capacities.}} 
	In all cases, the adversary (Eve)   
	has access to all  
	public messages,  denoted by $\vect F$. In addition to  $\vect F$, Eve might have  
	side information about $X_{\mc M}^n$. 
	When the adversary has no side information about $X_{\mc M}^n$, then $Z^n={\text{constant}}$ (i.e., independent of $X_{\mc M}^n$),  and  
	the capacity is called {\em SK capacity} and is denoted by $C_{SK}^{\mc A}(Q)$. 
	The adversary may compromise a subset of terminals $\mc D \subset \mc A^c$, in which case  
	Eve's side infromation is of the form $Z^n=X_{\mc D}^n=(X_j^n|~\forall j\in\mc D)$.
	The compromised terminals are  
	cooperative in the SKA protocol (e.g., they can be required by the SKA protocol to reveal $X_{\mc D}^n$ to other terminals.)  
	The capacity for this case is called {\em PK capacity }and is denoted by $C_{PK}^{\mc A|\mc D}(Q)$. In the most general sense, if Eve has access to side information $Z^n$, which is correlated with $X_{\mc M}^n$, the key capacity  is called {\em WSK (wiretap secret key) capacity} and is denoted by $C_{WSK}^{\mc A}(Q)$. Note that these key capacity definitions refer to a source model capacity if $Q$ is a joint distribution, and to a channel model capacity, if $Q$ is a conditional distribution.

	\section{General Lower and Upper Bounds}\label{sec:up_and_low}
	
	In this section, we give general lower and upper bounds for the SK and PK capacities of the channel model of transceivers. 
	Later, in Section~\ref{sec:nonint_cap}, we give a tightness condition under which these lower  and upper bounds 
	are matching. 
	
	\subsection{Lower Bound}\label{sec:lower}

	Before stating our main lower bound result we recall  the  single-letter characterization of the general source
	model PK capacity as given in \cite{Csiszar2004a}. 
	
	\begin{theorem}[PK Capacity \cite{Csiszar2004a}]\label{thm:PK-Cap-CN04}
		In a given source model ${X_\mc{M}}$ described by $P_{X_\mc{M}}$, for sharing a secret key among terminals in $\mc{A}\subsetneq \mc{M}$, with compromised terminals $\mc{D}\subseteq \mc{A}^c$, the PK capacity is
		\begin{IEEEeqnarray}{rCl}\label{eq:PK}
			C_{PK}^{\mc A|\mc D}(P_{X_\mc{M}}) &=& H(P_{X_{\mc{M}}}|P_{X_{\mc{D}}}) - R_{CO}^{\mc A|\mc D}(P_{X_\mc{M}}),\IEEEeqnarraynumspace 
		\end{IEEEeqnarray} where ${R_{CO}^{\mc A|\mc D}(P_{X_\mc{M}})= \min\limits_{R_{\mc{D}^c}\in\mc{R}_{CO}}\myop{sum}(R_{\mc{D}^c})}$ 
		and $\mc{R}_{CO} = 
		\left\{ R_{\mc{D}^c} | \myop{sum}(R_\mc{B}) \geq  H(P_{X_{\mc{M}}}|P_{X_{\mc{B}^c}}),~\forall \mc{B}\subset \mc{D}^c, \mc{A}\nsubseteq \mc{B}  \right\}.$
	\end{theorem}

	Equation~\eqref{eq:PK}  
	implies the SK capacity when ${\mc D = \emptyset}$.  
	The achievability result is based on a protocol  in which first,  
	the compromised terminals (that are assumed to be cooperative) reveal their observed random variables, and then the rest of the terminals in $\mc D^c$ communicate over the public channel to attain omniscience (i.e., the state that terminals in $\mc D^c$ learn each other's initial observations). Finally,  
	terminals in $\mc A$ extract the key from the common shared randomness $X^n_{\mc M}$.  
	It was also showed that the public communication required to obtain this PK capacity can be noninteractive, meaning that $\vect F=\vect F^n = (F_1,\ldots , F_m)$, where $F_j = X_j^n$ for all $j\in\mc D$ and $F_j = f(X_{\mc D}^n, X_j^n)$ for all $j\in\mc D^c$. 
	See the achievablity part of the proof of Theorem 2, in Section IV of \cite{Csiszar2004a}.

	We next review the \emph{source emulation} approach of \cite{Csiszar2008,Csiszar2013}. 
	Consider the multiaccess channel model (see Section \ref{sec:multiaccess}).  
	The \emph{simple source emulation}, introduced in \cite{Csiszar2008}, works as follows. 
	For a known IID input distribution $P_{X_{\mc I}}$, 
	each input terminal $j\in\mc I$ samples IID  symbols $X^n_j$ and transmits their symbols through the DMC. 
	During these $n$ symbol transmissions, terminals do not engage in public discussion. 
	After the  symbol transmissions, all terminals have $n$ IID samples according to the IID distribution given by 
	$P_{X_{\mc M}} = P_{X_{\mc I}} P_{X_{{\mc{M} \setminus \mc I}}| X_{{\mc{I}}}}$. 
	This way, in effect, a source model with a known IID distribution is realized (or \emph{emulated}) among terminals of $\mc M$. 
	Thus, after the symbol transmission steps, any suitable source model SKA protocol can be utilized for key generation.  
	
	The source emulation technique is proved \cite{Csiszar2008} to be capacity achieving for single-input multiple-out channels-- i.e., when $k=1$ and $\mc I=\{1\}$. However, in general, using public discussion during  symbol transmission can potentially result in more powerful and tighter lower bounds for the multiaccess channel model. This was proved in affirmative for some special multiaccess channels in \cite[Theorem 4]{Tyagi2013d}.  
	
	We use the \emph{general source emulation} approach  
	and prove the following theorem.
	
	\begin{theorem}\label{thm:lower}
		For a  
		channel model of $m$ transceivers $P_{Y_{{\mc{M}}}|T_{{\mc{M}}}}$, with $\mc M=[m]$,  and for any publicly known random variable $V$ satisfying $P_{V,T_{{\mc{M}}}} = P_V \Pi_{j\in {\mc{M}}} P_{T_j|V}$, we have
		\begin{equation*}
		C_{SK}^{\mc A}(P_{Y_{{\mc{M}}}|T_{{\mc{M}}}}) \geq C_{PK}^{\mc A|\{0\}}(P_{V T_{{\mc{M}}}} P_{Y_{{\mc{M}}}|T_{{\mc{M}}}}),
		\end{equation*}
		and 
		\begin{equation*}
		C_{PK}^{\mc A|\mc D}(P_{Y_{{\mc{M}}}|T_{{\mc{M}}}}) \geq C_{PK}^{\mc A|{\mc D'}}(P_{V T_{{\mc{M}}}} P_{Y_{{\mc{M}}}|T_{{\mc{M}}}}),
		\end{equation*}
		where $C_{PK}^{\mc A|{\mc D'}}(P_{V T_{{\mc{M}}}} P_{Y_{{\mc{M}}}|T_{{\mc{M}}}})$ denotes the  
		emulated source model PK capacity of  an auxiliary model with $m+1$ terminals, defined over
		${\mc M'} = \{0,1,\ldots,m\}$, where ${\mc D'} = \mc D \cup \{0\}$, $X_0 = V$, $X_j=X_j~\forall j\neq 0$, and an  
		underlying 
		source distribution 
		$P_{X_{{\mc M '}}} = P_V (\Pi_{j\in\mc M} P_{T_j|V}) P_{Y_{\mc M}|T_{\mc M}}$. 
	\end{theorem}

		\begin{IEEEproof}
			We show that for a transceiver channel model for terminal set $\mc M$, one can construct a source model for terminal set $\mc M'= \{0\}\cup\mc M$, and use source model protocols of \cite{Csiszar2004a} in the latter model to obtain a channel model SKA protocol in the transceiver model. 
			This leads to a lower bound on the PK capacity of the transceiver model. 
			The case of SK capacity is implied from the argument with $\mc D=\emptyset$.

			For a given transceiver channel model $P_{Y_{\mc M}|T_{\mc M}}$ for terminal set $\mc M$ 
			define an associated  source  model $P_{X_{{\mc M'}}}$  defined over $\mc M'= \{0\}\cup\mc M$,
			where $\{0\}$ is a new terminal added to the terminal set. 
			Let $V$ denote the random variable of terminal $0$. 
			The distribution  of this source model is given by $P_{X_{{\mc M '}}} = P_V (\Pi_{j\in\mc M} P_{T_j|V}) P_{Y_{\mc M}|T_{\mc M}}$,
			where $P_V$ and $P_{T_j|V}$ are arbitrary distributions that together generate a distribution on the input symbols of the transceivers channel. 
			Thus the distribution of $P_{T_{\mc M}}$ can be viewed as obtained from symbol transmission over a single-input multi-output channel, 
			where $V$ is the input symbol, and $T_{\mc M}$ denotes output symbols. 
			See Figure~\ref{fig:UpperBoundAuxChMod}~(a). 
			Note that $T_j$'s are assumed conditionally independent given $V$, that is $P_{T_{\mc M}|V} = \Pi_{j} P_{T_j|V}$.   
			Terminal $0$ is assumed compromised, hence $\mc D' =\{0\}\cup\mc D$.

			Let $K$ be a secret key  generated for terminals in  $\mc A$ by the noninteractive protocol $\Pi$ that was outlined  
			in the discussion after Theorem~\ref{thm:PK-Cap-CN04}, and achieves the {\em source model} PK capacity $C_{PK}^{\mc A|{\mc D'}}(P_{X_{{\mc M'}}})$. 
			In $\Pi$, the public message of terminal $j$ is a function of $T_j^n$ and $Y_j^n$.  
			The key $K$ is a function of $X^n_{{\mc M'}}$ and $\vect F$.  
			The protocol $\Pi$ defines a protocol $\Pi'$ for the transceivers  model,  using the following steps. 
			First, we emulate (realize) the source model $P_{X_{{\mc M'}}}$.  
			Note that
			$P_{V T_{\mc M}}$ is known. 
			Let $v^n=(v_1,\ldots,v_n)$ be a realization of $V^n$, accessible to all terminals and 
			publicly known. 
			Each input symbol $T_j^n$ is generated independently (given $v^n$) according to $P_{(T_j)t}=P_{T_j|V=v_t}$ for all $t\in[n]$. 
			In $n$ consecutive rounds, terminals use the DMC $P_{Y_{\mc M}|T_{\mc M}}$, without using the public communication channel. 
			Thus, after symbol transmission, source model $P_{X^n_{{\mc M'}}}$ is emulated for terminals in $\mc M$. That is each terminal $j$  has access to IID random variables  $T_{j}^n$ and $Y_{j}^n$ distributed according to the source distribution  $P_{X^n_{{\mc M'}}}$. 
			Now, terminals in $\mc M$ can run the source model SKA $\Pi$. 
			Compromised terminals send their samples $X_{\mc D}^n$ over the public channel. 
			The samples of  terminal $0$ is also accessible to the rest of the terminals. 
			Messages of terminals in $\mc M\setminus \mc D'$ are generated according to $\Pi$. 
			Then, all terminals in $\mc A$ can agree on the common randomness $X_{\mc M'}$, and extract their secret key. 
			Thus, at the end of $\Pi'$ the same key $K$ of $\Pi$ will be established for $\mc A$, 
			and $\Pi'$ provides a lower bound on the PK capacity of the transceiver channel model. 
			The key rate of $\Pi'$ is the same as the key rate of $\Pi$ which can be as large as $C_{PK}^{\mc A|{\mc D'}}(P_{X_{{\mc M'}}})$. 
		\end{IEEEproof}

	\subsection{Upper Bound}\label{sec:upper}
	
\begin{figure}[t]
		\centering
		\includegraphics[width=0.9\linewidth]{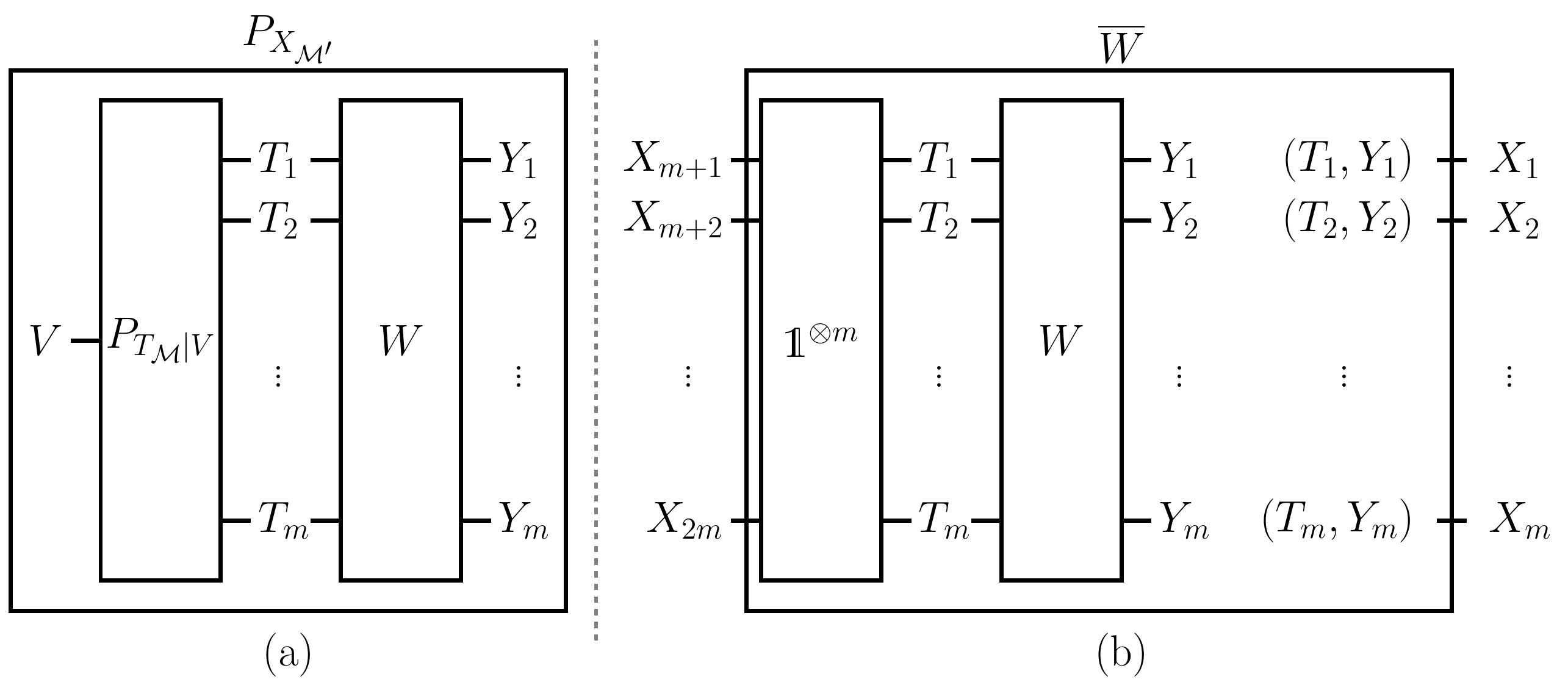}
		\caption{(a) The auxiliary source model $P_{X_{{\mc M}'}}$ used in the lower bound of Theorem~\ref{thm:lower}. (b) The auxiliary multiaccess channel model $\overline{W}$ used in the upper bound of Theorem~\ref{thm:upper_gen}.}
		\label{fig:UpperBoundAuxChMod}
	\end{figure}

	We demonstrate a new relation  between the proposed channel model of transceivers and the multiaccess channel model of \cite{Csiszar2013} to prove an upper bound on the SK and PK capacities of any given channel model of transceivers. 
	For a  
	transceiver channel model we define an auxiliary (related) multiaccess channel model, which is explicitly defined in  
	the following (also see Figure~\ref{fig:UpperBoundAuxChMod}~(b)). 
	Consider a given general channel model of transceivers $W$ defined by $P_{Y_{\mc M}|T_{\mc M}}$. Given $P_{Y_{\mc M}|T_{\mc M}}$ over terminal set $\mc M$, define an auxiliary multiaccess channel model $\overline W$ over $2m$ terminals denoted by $\overline{\mc M}$. 
	Let $\mc M' = \{m+1, \ldots,  2m\}$ be the set of new  input terminals. Let $\mc M$ be the set of output terminal of  $\overline W$ and thus, $\overline{\mc M} = \{1,\ldots , m, m+1, \ldots 2m\} =  \mc M' \cup \mc M$. 
	Input terminals have the special property that  $X_j=T_{j-m} ~\forall j \in \mc M' = \{m+1, \ldots,  2m\}$, and output terminals are defined as per the given transceivers model, i.e., $X_j = (T_j , Y_j) ~\forall j\in \mc M = \{1, \ldots,  m\}$. 
	The multiaccess channel model $\overline W$ can be viewed as a two-layer DMC where the first layer is a DMC with input RVs $X_{m+1},\ldots, X_{2m}$, and output RVs $X_1, \ldots, X_m$ such that $P_{T_{\mc M}|X_{\mc M'}} = \Pi_{j\in\mc M} P_{T_j|X_{j+m}}$ and $P_{T_j|X_{j+m}} = \mathds{1}(T_j=X_{j+m})$ is an identity (noiseless) channel for all $j\in\mc M$. The DMC of the second layer is given by $P_{Y_{\mc M}|T_{\mc M}}$.

	\begin{theorem} \label{thm:upper_gen}
		The channel model SK capacity and the channel model PK capacity of $P_{Y_{\mc M}|T_{\mc M}}$ for any $\mc D\subset \mc M=[m]$, and $\mc A\subseteq \mc D^c$ are upper bounded by
		\begin{align}
		C_{SK}^{\mc A}(P_{Y_{\mc M}|T_{\mc M}}) &\leq C_{SK}^{\mc A}(P_{T_{\mc M}|X_{\mc M'}} P_{Y_{\mc M}|T_{\mc M}}), 
		\end{align}
		and 
		\begin{align}
		C_{PK}^{\mc A|\mc D}(P_{Y_{\mc M}|T_{\mc M}}) &\leq C_{PK}^{\mc A|\mc D}(P_{T_{\mc M}|X_{\mc M'}} P_{Y_{\mc M}|T_{\mc M}}), 
		\end{align}
		where the right hand side of the above inequalities are the SK and PK capacities of the auxiliary model $\overline W$.
	\end{theorem}

		\begin{IEEEproof}
			We show that the PK capacity of $W$ gives an achievable lower bound for $\overline{W}$. 
			Let $K\in\mc K$ be a private key established for $W$ by SKA protocol $\Pi$ 
			such that $\log |\mc K| \leq n C_{PK}^{\mc A|\mc D}(P_{Y_{\mc M}|T_{\mc M}})$.  
			We use $\Pi$ to generate a key $K'\in\mc K$ in $\overline{W}$.   
			First note that in the auxiliary multiaccess channel model of $\overline{W}$, 
			after each symbol transmission each 
			terminal $j\in\mc M$ has access to the same variable(s) of the input terminal $j+m\in\mc M'$. 
			Therefore, terminals of $\mc M'$ can always remain silent (not sending public messages), 
			and all public messages can be generated by terminals in $\mc M$, the output terminals of $\overline{W}$.  
			Thus, helper terminals of $\mc M'$ are dummy terminals, and their presence can only help with the key generation. 
			Let $\Pi$ be such that in each round $t \leq n$, terminals generate and send input symbols $T_{jt} = \tilde T_{jt}$'s to $W$ and 
			receive corresponding output symbols $Y_{jt} = \tilde Y_{jt}$.   
			Then terminals engage in a public discussion $\tilde{\vect F}^t$. 
			Let $\Pi'$ be the protocol for SKA in $\overline{W}$ which works as follows. 
			In each round $t$, input terminals  $j+m\in\mc M'$ generate and send input symbols $X_{(j+m)t} = \tilde T_{jt}$'s to  $\overline{W}$. 
			Note that for every terminal $j\in\mc M$, we have $T_{jt} = \tilde T_{jt}$ and $Y_{jt} = \tilde Y_{jt}$.  
			Then, input terminals  $j+m\in\mc M'$ remain silent 
			and output terminals of multiaccess channel $\overline{W}$  in $\mc M$ invoke  public discussion $\vect F^t = \tilde{\vect F}^t$.  
			Following the same instructions of $\Pi$, at the end of round $n$,  
			terminals in $\mc M$ can agree on a secret key $K'\in \mc K$. 
			As, in effect, $\Pi$ and $\Pi'$ are identical protocols from the view point of $\mc M$, $K'$ is equal to $K$. 
			Therefore,  $(1/n) \log |\mc K|$  is also an achievable key rate for the multiaccess model of $\overline{W}$.  
			The maximum rate of such key is given by the PK capacity of ${W}$.  
			The argument for SK capacity is the same with $\mc D=\emptyset$. 
		\end{IEEEproof}

	Theorem \ref{thm:upper_gen} implies that  an  
	upper bound for the SK (or PK) capacity of $\overline{W}$, including the upper bounds of \cite[Theorem 6]{Csiszar2013}, is an upper bound for $C_{SK}^{\mc A}(P_{Y_{\mc M}|T_{\mc M}})$ (or $C_{PK}^{\mc A|\mc D}(P_{Y_{\mc M}|T_{\mc M}})$).

	\section{The Noninteractive SK Capacity}\label{sec:nonint_cap}

	We consider the following special case of the  transceiver channel model (defined in Section~\ref{sec:defs_model}).
	
	\textbf{(a) Noninteractive SKA.} The public communication $\vect F$ is noninteractive. This means that $\vect F$ only occurs after $n$ rounds of symbol transmission and each terminal  sends one public message. Thus, $X_{\mc M}^n$ is not a function of $\vect F$. 
	
	\textbf{(b) Independent Inputs.} Terminals are locally controlling their input variables, and  
	the input variables are independent, i.e., $P_{T_{\mc M}}=\Pi_{j\in\mc M} P_{T_{j}}$. 
	
	Key agreement protocols with such properties are  
	desirable in practice as they use less resources and  are easier to realize.  
	
	Let $C_{NI-SK}^{\mc A}(P_{Y_{{\mc{M}}}|T_{{\mc{M}}}})$ denote the largest achievable key rate of all noninteractive SKAs that satisfy  
	conditions (a) and (b), above. 
	The noninteractive SK capacity has been  
	extensively  studied in the two-party model; see for example \cite{Ahlswede1993,Holenstein2005,Renes2013}.

	\begin{theorem}\label{thm:nonint-cap}
		For a  
		channel model of transceivers,  and a  
		subset $\mc A\subseteq\mc M$, the noninteractive SK capacity is given by
		\begin{equation}
		C_{NI-SK}^{\mc A}(P_{Y_{\mc M}|T_{\mc M}}) = \max_{P_{T_{\mc M}}} C_{SK}^{\mc A}(P_{T_{\mc M}} P_{Y_{\mc M}|T_{\mc M}}).
		\end{equation}
	\end{theorem}
	
	\begin{IEEEproof}
		First, we prove that the right hand side of the above equation is an upper bound on the noninteractive capacity.  Consider an auxiliary multiaccess channel $\overline{W}$ with $2m$ terminals denoted by $\overline{\mc M} = \{1,\ldots , m, m+1, \ldots 2m\}$, where $X_j = (T_j , Y_j) ~\forall j\in \mc M = \{1, \ldots,  m\}$, are output variables of the multiaccess DMC and $X_j~\forall j \in \mc M' = \{m+1, \ldots,  2m\}$ are input variables of DMC, satisfying $P_{T_{\mc M}|X_{\mc M'}} = \Pi_{j\in[m]} P_{T_j|X_{j+m}}$ and $P_{T_j|X_{j+m}} = \mathds{1}(T_j=X_{j+m})$.  
		By Theorem \ref{thm:upper_gen}, we have 
		\begin{equation*}
		C_{NI-SK}^{\mc A}(P_{Y_{\mc M}|T_{\mc M}}) \leq C_{NI-SK}^{\mc A}(P_{T_{\mc M}|X_{\mc M'}}P_{Y_{\mc M}|T_{\mc M}}),
		\end{equation*}
		and thus any upper bound on 
		the noninteractive SK capacity of multiaccess model $\overline{W}$ is also an upper bound on
		the noninteractive SK capacity of the transceiver model $W$. An upper bound is given for the SK capacity of any  general multiaccess channel model in \cite{Csiszar2013}. Before using the proof of that upper bound we  
		define the following notations.  
		
		For  a  
		subset $\mc A\subset \mc M$, let $\Gamma(\mc A)$ be the family of all nonempty sets $\mc B \subset \mc M$ such that,
		$$\mc A \nsubseteq \mc B, \quad\forall \mc B\in\Gamma(\mc A),$$ and let $\Lambda(\mc A)$ be the set of all $|\Gamma(\mc A)|$-dimensional vectors $\lambda = \{\lambda_{\mc B}~:~ \mc B\in\Gamma(\mc A) \}$ that satisfy the following two conditions:
		\begin{align*}
		0\leq \lambda_{\mc B}\leq 1, ~\quad \forall \lambda\in\Lambda(\mc A),\\
		\sum_{\mc B\in \Gamma(\mc A)~:~ j\in \mc B} \lambda_{\mc B} = 1,~\quad \forall j\in\mc D^c \text{~and~} \forall \lambda.
		\end{align*}
		According to the proof of Theorem 6, Eq.~(28), in \cite{Csiszar2013}, for the multiaccess channel $\overline{W}$, where $\mc M'$ is the set of input terminals (transmitters), $\mc M$ is the set of output terminals (receivers), and $\mc D=\emptyset$, for any $\lambda$, any achievable key $K$ satisfies 
		\begin{align}\label{eq:En}
		\frac{1}{n}\log|\mc K|\leq \frac{\alpha_n}{n}E_n + \beta_n,
		\end{align}
		where $\alpha_n\to1$ and $\beta_n\to 0$, as $n\to\infty$; and 
		\begin{align*}
		E_n&=\sum_{t=1}^n \big[ \big(H(X_{\mc M t}) - \sum_{\mc B\in\Gamma(\mc A)} \lambda_{\mc B}H(X_{{\mc B}t}|X_{{\mc B^c}t}) \big)\\
		& \quad- \big(H(X_{\mc M' t}) - \sum_{\mc B\in\Gamma(\mc A)} \lambda_{\mc B}H(X_{({\mc B}\cap \mc M')t}|X_{({\mc B^c}\cap \mc M')t}) \big)\big].
		\end{align*}
		Note that under $n\to\infty$ the right hand side of \eqref{eq:En} gives an upper bound on the SK capacity of the auxiliary multiaccess model $\overline{W}$ which, because of Theorem~\ref{thm:upper_gen},  implies an upper bound on the SK capacity of $W$.  
		Now we use  
		assumptions (a) and (b) to simplify the expression of $E_n$ in \eqref{eq:En}. Due to noninteractivity assumption (a), we have  
		\begin{align*}
		E_n&=\big(H(X_{\mc M}^n) - \sum_{\mc B\in\Gamma(\mc A)} \lambda_{\mc B}H(X^n_{{\mc B}}|X^n_{{\mc B^c}}) \big)\\
		& \quad- \big(H(X^n_{\mc M'}) - \sum_{\mc B\in\Gamma(\mc A)} \lambda_{\mc B}H(X^n_{{\mc B}\cap \mc M'}|X^n_{{\mc B^c}\cap \mc M'}) \big).
		\end{align*}
		By the independence of the inputs assumption (b),   and  properties of $\lambda$ vectors,  we have $H(X_{\mc M'})=\sum_{j\in\mc M'} H(X_j)$ and  
		\begin{align*}
		E_n&=n\big(H(X_{\mc M}) - \sum_{\mc B\in\Gamma(\mc A)} \lambda_{\mc B}H(X_{{\mc B}}|X_{{\mc B^c}}) \big).
		\end{align*}
		By Theorem 3.1 of \cite{Csiszar2008} we know that for $Q=P_{T_{\mc M}} P_{Y_{\mc M}|T_{\mc M}}$ 
		\begin{align*}
		C_{SK}^{\mc A}(Q) = \min_{\lambda\in\Lambda(\mc A)} H(X_{\mc M}) - \sum_{\mc B\in\Gamma(\mc A)} \lambda_{\mc B}H(X_{{\mc B}}|X_{{\mc B^c}}),
		\end{align*}
		which completes the proof of the converse. Also note that, a simple noninteractive source emulation method of Theorem \ref{thm:lower} with $V=\text{constant}$ achieves this capacity. 
	\end{IEEEproof}
	\subsection{The case of Polytree-PIN}

		In  the following, we narrow our focus on a channel model of transceivers  that  
		can be described by a \emph{Polytree}, that is defined as follows. Assume that $G=(\mc M,\mc E)$ is a directed graph, where $\mc M$ is the set of vertexes and $\mc E$ is the set of directed edges. Each vertex (node) represents a terminal in the SKA model. Let $\mymtrx{UD}(G) \to G'=(\mc M, \mc E')$ is a mapping that for a given directed graph $G$ gives an undirected graph $G'$ over the same nodes $\mc M$ such that if in $G$, nodes $i$ and $j$ are connected by a directed edge $e_{ij}\in\mc E$, nodes  $i$ and $j$ are  also connected by an undirected edge $e_{ij}\in\mc E'$ in $G'$. For any directed graph $G$, $\mymtrx{UD}(G)$ is called the the underlying  undirected graph of $G$.  
		We call  $G$ a polytree, if its underlying  undirected graph is a tree (has no cycles) and  
		say a channel model is a polytree-PIN model if,  
		$T_i = (T_{ij}| ~j\in {\mc{M}}\setminus\{i\})$, $Y_i = (Y_{ij}| ~j\in {\mc{M}}\setminus\{i\})$, and 
		$$P_{Y_{{\mc{M}}}|T_{{\mc{M}}}}= \prod_{e_{ij}\in \mc E} P_{Y_{ij}|T_{ji}},$$ where $e_{ij}$ denotes a directed edge of the polytree $G$. 
		The Polytree-PIN model is basically the channel model counterpart of Tree-PIN model of \cite{Poostindouz2019}, that is a special class of pairwise independent network (PIN) source models \cite{Ye2007,Nitinawarat2010c}. PIN models are of special interest 
    	because they model  
    	correlations that are generated through the application of wireless communication networks. 
	
	\begin{corollary}
		For any given Polytree-PIN defined by $G=(\mc M,\mc E)$ and probability distribution $P_{Y_{{\mc{M}}}|T_{{\mc{M}}}}$, the noninteractive channel model SK capacity is given by
		\begin{align*}
		C_{NI-SK}^{\mc A}(P_{Y_{{\mc{M}}}|T_{{\mc{M}}}}) = \max_{P_{T_{{\mc{M}}}}} \min_{\substack{i,j\in\mc M \\ \textrm{~s.t.~} e_{ij}\in\mc E}} I(T_{ij};Y_{ji}).
		\end{align*}
	\end{corollary}
	
	The proof is due to Theorem~\ref{thm:nonint-cap},  and the fact that the  source model $P_{T_{\mc M}}P_{Y_{\mc M}|T_{\mc M}}$ is a Tree-PIN for which the SK capacity is equal to $\min_{i,j} I(T_{ij};Y_{ij})$ (See \cite{Poostindouz2020} Theorem 2, and Eq.(36) of \cite{Csiszar2004a}.)

	\subsection{The case of wiretapped Polytree-PIN}

	In practice, there are many cases where the adversary is powerful and is capable of wiretapping. The first information theoretical treatment of a two-party wiretapping scenario was considered for secure message transfer by Wyner in \cite{Wyner1975}. In the context of key agreement also it is always desirable to consider scenarios where the adversary has access to some wiretapped side information. Some wiretapped models  for SKA and their corresponding wiretap secret key (WSK) capacities were studied in \cite{Ahlswede1993,Maurer1993,Csiszar2008}; however, these models are limited in comparison to the general models considers for SK and PK capacities.

	Unfortunately, the WSK capacity of the general source model as defined previously, even for the special case of two terminals (${|\mc M|=2}$) remains an open problem. For the case of two-party SKA, the source model WSK capacity is upper bounded by  $I(X_1;X_2|Z)$, which is proved to be a tight bound under the additional assumption that the Markov Chain $X_1 - X_2 - Z$ holds \cite{Ahlswede1993,Maurer1993}.   
	In the case of $ |\mc M|\geq 3$, PK capacities (e.g., Theorem \ref{thm:PK-Cap-CN04}) lead to an upper bound for the WSK capacity. 
		
		\begin{lemma}
			For a given general wiretapped (source or channel) model $Q$, let $C_{PK}^{\mc A|\{m+1\}}(Q')$ be the PK capacity of a model with $m+1$ terminals such that $X_j=X_j$ for all $j\leq m$, and $X_{m+1} = Z$, where only terminal $m+1$ is compromised (i.e., $\mc D = \{m+1\}$). By definition of the PK capacity we have $C_{WSK}^{\mc A}(Q) \leq C_{PK}^{\mc A|\{m+1\}}(Q').$ 
		\end{lemma}
		
		\begin{IEEEproof}
			Knowledge of $Z^n$ by terminals in $\mc A$ can only be used to help with secret key extraction (privacy amplification). Any achievable $(\eps,\sigma)-$SK for model $Q$ is also an achievable $(\eps,\sigma)-$SK for model $Q'$, thus the WSK capacity of $Q$ is a lower bound to the PK capacity of $Q'$.  
		\end{IEEEproof}

	We call a model, wiretapped polytree-PIN, if it is a polytree-PIN channel model with $G=(\mc M,\mc E)$ and for any $e_{ij}\in\mc E$ the Markov relation $T_{ij}-Y_{ji}-Z_{ij}$ holds, where $Z=(Z_{ij}| i,j\in\mc M, \text{s.t.~} e_{ij}\in\mc E)$ is the wiretapper side information. 
	
	\begin{theorem} 
		The channel model WSK capacity of any given wiretapped polytree-PIN is lower bounded by
		\begin{align} \nonumber
		C_{WSK}^{\mc A}(P_{ZY_{{\mc{M}}}|T_{{\mc{M}}}}) &\geq \max_{P_{T_{{\mc{M}}}}} C_{WSK}^{\mc A}(P_{T_{{\mc{M}}}}P_{ZY_{{\mc{M}}}|T_{{\mc{M}}}}) \\
		&= \max_{P_{T_{{\mc{M}}}}} \min_{\substack{i,j\in\mc M \\ \textrm{~s.t.~} e_{ij}\in\mc E}} I(T_{ij};Y_{ji}|Z_{ij}).
		\end{align}
	\end{theorem}
	
	\begin{IEEEproof}
		For any given $P_{T_{\mc M}}$, any secret key $K$ achieving the wiretap secret key (WSK) capacity of the source model described by $P_{T_{{\mc{M}}}}P_{ZY_{{\mc{M}}}|T_{{\mc{M}}}}$ is also a wiretap secret key for the wiretapped channel model of $P_{ZY_{{\mc{M}}}|T_{{\mc{M}}}}$. The source model WSK capacity is given in \cite{Poostindouz2020} Theorem 2. Thus, for any $P_{T_{\mc M}}$, the emulated source model WSK capacity is lower bound to the channel model WSK capacity of $P_{ZY_{{\mc{M}}}|T_{{\mc{M}}}}$. A maximization over $P_{T_{\mc M}}$ gives the best of such lower bounds. 
	\end{IEEEproof}

	\begin{theorem} 
		The channel model WSK capacity of any given wiretapped polytree-PIN for any $\mc A\subseteq \mc M=[m]$, is upper bounded by
		\begin{align}
		C_{WSK}^{\mc A}(P_{ZY_{\mc M}|T_{\mc M}}) &\leq C_{PK}^{\mc A|\{2m+1\}}(P_{T_{\mc M}|X_{\mc M'}} P_{ZY_{\mc M}|T_{\mc M}}) 
		\end{align}
		where the right hand side of the above inequality is the PK capacity of $\overline W$, an auxiliary multiaccess channel model with $2m+1$ terminals $\overline{\mc M} = \{1,\ldots , m, m+1, \ldots 2m, 2m+1\}$, where $\mc D = \{2m+1\}$ is the compromised terminal (i.e., $X_{\mc D}=X_{2m+1} = Z$), $X_j = (T_j , Y_j) ~\forall j\in \mc M = \{1, \ldots,  m\}$, $X_j=T_j ~\forall j \in \mc M' = \{m+1, \ldots,  2m\}$, and $\overline W$ is a degraded multiaccess channel with input RVs $X_{m+1},\ldots, X_{2m}$, and output RVs $X_1, \ldots, X_m,Z$ such that $P_{T_{\mc M}|X_{\mc M'}} = \Pi_{j\in{\mc{M}}} P_{T_j|X_{j+m}}$ and $P_{T_j|X_{j+m}} = \mathds{1}(T_j=X_{j+m})$.
	\end{theorem}
	
	The proof is in the same lines as for the proof of Theorem~\ref{thm:upper_gen}. 
	Any upper bound to the aforementioned multiaccess channel model $\overline{W}$ is therefore also an upper bound for $C_{WSK}^{\mc A}(P_{ZY_{\mc M}|T_{\mc M}})$.

	\section{Conclusion}\label{sec:conclusion}
	
	We introduced a new general channel model for multiterminal secret key agreement. Channel models of \cite{Csiszar2008} and \cite{Csiszar2013} are shown to be the special cases of our proposed model of transceivers. We gave  lower bounds and upper bounds for the SK and PK capacities of the transceivers model. Then, we studied the problem of noninteractive secret key agreement and gave the noninteractive SK capacity of the transceivers model. We gave a simpler expression for calculating the noninteractive SK capacity of Polytree-PIN as an example. Finding tighter bounds for the SK and PK capacity of the general case and investigating interactive protocols for obtaining tighter bounds for the case wiretapped Polytree-PIN are interesting research directions we leave for future work.

	\section{Acknowledgment} 
	This research is in part supported by Natural Sciences and Engineering Research Council of Canada, Discovery Grant program.


\end{document}